\documentstyle[preprint,aps,prb]{revtex}
\begin{document}
\draft
%\twocolumn[\hsize\textwidth\columnwidth\hsize\csname @twocolumnfalse\endcsname
\title{Single hole dynamics in dimerized spin liquids}
\author{O.P. Sushkov}
\address{School of Physics, University of New South Wales, Sydney 2052, Australia}

%\date{\today}
\maketitle

\begin{abstract}
 The  dynamics of a single hole in quantum antiferromagnets 
is influenced by  magnetic fluctuations.
 In the present
work we consider two situations. The first one corresponds to a single hole in 
the two leg $t-J$ spin ladder. In this case the wave function renormalization
is relatively small and the quasiparticle residue of the  $S=1/2$ 
state remains close 
to unity. However at large $t/J$ there are higher spin ($S=3/2,5/2,..$)
bound states of the hole with the magnetic excitations,
 and therefore there is a
crossover from  quasiparticles with $S=1/2$ to  quasiparticles with
higher spin.

The second situation corresponds to a single hole in two coupled 
antiferromagnetic planes very close to the point of antiferromagnetic
instability. In this case the hole wave function renormalization is very
strong and the quasiparticle residue vanishes at the point of instability.
\end{abstract}


\section{Introduction}
Quantum spin ladders  have recently become a subject of considerable interest,
mainly because of their relevance to 
a variety of quasi one-dimensional materials \cite{Dagotto}.
In addition  they represent, on the theoretical side, an important step  
towards 
understanding of the  two-dimensional CuO$_2$ planes 
in the  cuprate superconductors.
The  properties of a single hole in the two leg ladder were investigated
numerically by Troyer, Tsunetsugu, and Rice \cite{Tro} and Haas and 
Dagotto \cite{Haas} using exact diagonalizations. More recently
Eder \cite{Eder} approached the same problem using both exact 
diagonalization and the Self Consistent Born Approximation.
Two most interesting properties have been observed in these works:
The first one is the  shift of the hole dispersion minimum to the point 
close to $k \approx \pi/2$. The second property is the closeness of the
quasiparticle residue to unity even at large $t/J$. Such behavior of the
residue is very much different from the $t-J$ model on the 2D square lattice
\cite{Dag1}.

The closeness of the quasiparticle residue to unity indicates that a simple
variational approach and even perturbation theory must be applicable to the
problem, and therefore we use these techniques in the present work.
This allows us to find the $S=1/2$ quasiparticle dispersion analytically.
The dispersion is in an agreement with previous computations.
We demonstrate also that there are higher spin ($S=3/2,5/2,...$)
bound states of the hole with the magnetic excitations.
At large $t/J$ these bound states lie below the  $S=1/2$ state, and 
this means that
they become the  real quasiparticles of the system.
So our conclusion is that the spectrum of the single hole states on the
ladder is very rich. This complexity is somewhat similar to the complexity
of the spin-wave spectrum \cite{Sus,Kotov}.

In spite of the complex spectrum, the structure of each individual state
on the ladder is relatively simple. It means that the number of significant
components in the wave function remains of the order of unity. The purpose 
of the
second part of this work is to consider a spin liquid with really complex 
structure of the hole wave function. We demonstrate that this situation is 
realized in a 2D spin liquid close to the point of 
 antiferromagnetic instability.
Specifically we consider two coupled antiferromagnetic planes. As
 the point of instability is approached the number of components in the hole
wave function  approaches  infinity and the quasiparticle residue
vanishes.

\section{ $t-J$ ladder}
We consider the standard $t-J$ model on a two-leg ladder. The Hamiltonian
reads
\begin{eqnarray}
\label{H}
&&H=H_t+H_J,\\
&&H_t=-\sum_{\langle i,j \rangle \sigma}
\left(t \ c^{\dag}_{i \sigma} c_{j \sigma}+H.c. \right),\nonumber\\
&&H_J=\sum_{\langle i,j \rangle} J_{ij} \ {\bf S}_i \cdot {\bf S}_j,\nonumber 
\end{eqnarray}
where $c_{i\sigma}^{\dag}$ is the creation operator of an electron
with spin $\sigma$ at site $i$. As usual  double occupancy is
forbidden. $\langle ij\rangle$ represents nearest neighbor sites.
The spin operator is ${\bf S}_i={1\over2}c_{i\alpha}^{\dag}
\sigma_{\alpha\beta}c_{i\beta}$. For bonds along the legs we choose
$J_{ij}=J$, and for bonds along the rungs $J_{ij}=J_{\perp}$.
We are mostly interested in the ``isotropic'' case: $J=J_{\perp}$,
but it is convenient to keep $J_{\perp}\ge J$ as an independent parameter.
Following Chubukov \cite{Chub}, Sachdev and Bhatt \cite{Sach}
and Gopalan, Rice, and Sigrist \cite{Copalan}
we consider the  ground state consisting of inter-chain  spin singlets
$|GS\rangle=|1,s\rangle |2,s\rangle |3,s\rangle ...$,
where $|n,s\rangle={1\over{\sqrt{2}}}\left[|\uparrow\rangle_i
|\downarrow\rangle_j-|\downarrow\rangle_i|\uparrow\rangle_j\right]$
as a starting point of our approach.
Since each  singlet can be excited into a  triplet state it is natural to 
introduce a creation operator $t_{\alpha n}^{\dag}$  for  this 
excitation:
\begin{equation} 
|n,\alpha \rangle=t_{\alpha n}^{\dag}|n,s\rangle, \ \ \alpha=x,y,z.
\end{equation}
The interaction of the triplets between themselves is 
 described by the Hamiltonian
(see, e.g. Refs. \cite{Copalan,Sus})
\begin{eqnarray}
\label{H1}
H_{sw}&=&\sum_{n,\alpha,\beta}\left\{J_{\perp}t_{\alpha n}^{\dag}t_{\alpha n}+
{{J}\over{2}}\left(t_{\alpha n}^{\dag}t_{\alpha n+1}+
t_{\alpha n}^{\dag}t_{\alpha n+1}^{\dag}+\mbox{h.c.}\right) \right.
\nonumber \\
& & \left. + {{J}\over{2}}\left(t_{\alpha n}^{\dag}
t_{\beta n+1}^{\dag}t_{\beta n}t_{\alpha n+1}
-t_{\alpha n}^{\dag}t_{\alpha n+1}^{\dag}t_{\beta n}t_{\beta n+1}\right)
\right\}.
%\end{equation}
\end{eqnarray}
In addition, the hard core constraint
\begin{equation}
\label{c1}  
t_{\alpha n}^{\dag}t_{\beta n}^{\dag}=0
\end{equation}
 has to be enforced on every
site \cite{Sach}. The  interactions (\ref{H1},\ref{c1}) lead to rearrangement
of the unperturbed ground state and nontrivial spectrum of the spin-wave
excitations \cite{K,Sus,Kotov}.

If we remove one electron from  rung $n$ a single hole state on this
rung is created. Let us denote by $a^{\dag}_{n,m,\sigma}$ the creation operator
of a hole with spin $\sigma$ on rung $n$, where the index $m=1,2$ represents
the legs of the ladder. This operator is defined by the equation
\begin{equation}
\label{a}
a^{\dag}_{n,m,\sigma}|n,s\rangle=c^{\dag}_{n,{\bar m},\sigma}|0\rangle,
\end{equation}
where ${\bar m}=2$ if $m=1$, and ${\bar m}=1$ if $m=2$. This is a  
fermionic operator which satisfies the 
 usual fermionic anticommutation relations.
The hole on the rung $n$ interacts with the 
triplet excitations on its nearest sites. This interaction can be easily
found by calculating all possible matrix elements of the initial
Hamiltonian (\ref{H}) (see Ref.\cite{Eder}). The result is
\begin{eqnarray}
\label{Hh}
H_h&=&-t\sum_{n,\sigma}\left(a^{\dag}_{n,2,\sigma} \ a_{n,1,\sigma}+H.c.\right)
+{t\over2}\sum_{n,m,\sigma}
\left(a^{\dag}_{n+1,m,\sigma} \ a_{n,m,\sigma}+H.c.\right)\\
&+&\sum_{n,\sigma,\pm}\left({\bf t}_n^{\dag}\left[
t \ {\bf S}_{n \pm 1,n}
+{J\over2} \ {\bf S}_{n \pm 1,n \pm 1}\right]+H.c. \right)\nonumber\\
&+&{t\over2}\sum_{n,m,\sigma}\left({\bf t}_{n+1}^{\dag}{\bf t}_n
a^{\dag}_{n,m,\sigma}a_{n+1,m,\sigma}+H.c.\right)\nonumber\\
&-&t\sum_n\left(i \ {\overline {\bf S}}_{n+1,n}
\left[{\bf t}_{n}^{\dag}\times {\bf t}_{n+1}\right] +H.c.\right)
+{J\over2}\sum_{n,\pm} \left(i \ {\overline {\bf S}}_{n,n}
\left[{\bf t}_{n \pm 1}^{\dag}\times {\bf t}_{n \pm 1}\right] +H.c.\right).
\nonumber
\end{eqnarray}
The index $\pm$ denotes  summation over nearest rungs.
Following Ref. \cite{Eder} we have introduced the notations
\begin{eqnarray}
\label{s}
&&{\bf S}_{l,n}={1\over2}\sum_{m,\alpha,\beta} 
(-1)^m a^{\dag}_{l,m,\alpha}{\vec \sigma}_{\alpha\beta}a_{n,m,\beta},\\
&&{\overline {\bf S}_{l,n}}={1\over2}\sum_{m,\alpha,\beta} 
a^{\dag}_{l,m,\alpha}{\vec \sigma}_{\alpha\beta}a_{n,m,\beta}.\nonumber
\end{eqnarray}
where ${\vec \sigma}$ is the vector of Pauli matrices.
In addition we have to impose the  constraint that a triplet excitation and
a hole can not coexist on the same rung
\begin{equation}
\label{c2}  
t_{\alpha n}^{\dag}a_{n,m,\sigma}^{\dag}=0.
\end{equation}

The effective Hamiltonian given by Eqs.(\ref{H1}) and (\ref{Hh})
($H_{eff}=H_{sw}+H_h$)  combined with the 
 constraints (\ref{c1}) and (\ref{c2}) represents an {\it exact}
mapping of the initial Hamiltonian (\ref{H}) in the sector with a single
hole \cite{many}. One can work with $H_{eff}$ by using standard 
 diagram techniques 
in  momentum representation. The constraint (\ref{c2}) can be taken
into account in a mean field fashion \cite{Eder} or by introducing a
fictitious (ghost) interaction similarly to the way it was done 
   for spin waves in the
paper \cite{K}. However fortunately the problem can be resolved in a
simpler way. Because of the sizable spin-wave gap there is no real long-range
dynamics on the ladder. Therefore the problem can be solved in
the coordinate representation by constructing a
 trial wave function that explicitly
obeys the constraints (\ref{c1}) and (\ref{c2}). 
First consider an isolated rung with a hole. Say this is the rung $n$.
The 
 energy of this state with respect to the undoped system is 
\begin{equation}
\label{e0}
\epsilon_0={3\over4}J_{\perp}+{3\over4}{{J^2}\over{J_{\perp}}}.
\end{equation}
The origin of the first term in $\epsilon_0$ is obvious:
one antiferromagnetic link is destroyed. The origin of the second term is
also simple. There are quantum fluctuations in the spin liquid state:
virtual excitations of the triplet pairs created  by the term 
$t_{\alpha n}^{\dag}t_{\alpha n+1}^{\dag}$ in the Hamiltonian (\ref{H1}).
This gives vacuum energy $3{{(J/2)^2}\over{-2J_{\perp}}}$ per link,
where the coefficient 3 is due to the number of possible 
polarizations \cite{ab}. The presence of a hole on  rung $n$ suppresses
fluctuations at the links $n-1,n$ and $n,n+1$ and this gives the second term 
in (\ref{e0}). Because of the hybridization along the rung the 
stationary 
states are symmetric and antisymmetric with hopping energies $-t$ and $+t$ 
respectively. Thus for an isolated rung one finds
\begin{eqnarray}
\label{rung}
a^{\dag}_{n,+,\sigma}&=&{{a^{\dag}_{n,1,\sigma}+a^{\dag}_{n,2,\sigma}}\over
{\sqrt{2}}}, \ \ \ 
\epsilon=\epsilon_0-t,\\
a^{\dag}_{n,-,\sigma}&=&{{a^{\dag}_{n,1,\sigma}-a^{\dag}_{n,2,\sigma}}\over
{\sqrt{2}}}, \ \ \ 
\epsilon=\epsilon_0+t.
\nonumber
\end{eqnarray}
Consider now a hole hopping from  rung $n$ to $n+1$:
$a^{\dag}_{n,+,\uparrow} \to a^{\dag}_{n+1,+,\uparrow}$.
Naively, according to the Hamiltonian (\ref{Hh}) the  effective matrix element
for this process is $t/2$. However this does not take into account quantum 
fluctuations of the spin-wave vacuum. By taking into account these
fluctuations we obtain the following  
initial wave function 
\begin{equation}
\label{in}
|i\rangle=\left[(1-{3\over2}\mu^2)+
\mu t^{\dag}_{\alpha,n-2}t^{\dag}_{\alpha,n-1}\right]a^{\dag}_{n,+,\uparrow}
\left[(1-3\mu^2)+
\mu t^{\dag}_{\beta,n+1}t^{\dag}_{\beta,n+1}
+\mu t^{\dag}_{\beta,n+2}t^{\dag}_{\beta,n+3}\right]|s\rangle.
\end{equation}
Here we explicitly present the part of the wave function corresponding to the
rungs from $n-2$ to $n+3$. The coefficient $\mu$ describes  the spin-wave
quantum fluctuations and according to the Hamiltonian (\ref{H1}) it is 
equal to
\begin{equation}
\label{mu}
\mu\approx-{J\over{4J_\perp}}
\end{equation}
The final state is
\begin{equation}
\label{fin}
|f\rangle=\left[(1-3\mu^2)+
\mu t^{\dag}_{\alpha,n-2}t^{\dag}_{\alpha,n-1}
+\mu t^{\dag}_{\alpha,n-1}t^{\dag}_{\alpha,n}
\right]a^{\dag}_{n+1,+,\uparrow}
\left[(1-{3\over2}\mu^2)
+\mu t^{\dag}_{\beta,n+2}t^{\dag}_{\beta,n+3}\right]|s\rangle.
\end{equation}
Calculating the matrix element of the Hamiltonian $H_t$, Eq.(\ref{H}) 
we find 
\begin{equation}
\label{hop}
t^{(1)}={t\over{2}}\rho^2, \ \ \ \rho=1-{3\over2}\mu^2,
\end{equation}
which agrees with that obtained in the Ref. \cite{Eder}.

Let us consider now the  possibility of  a virtual spin-wave excitation
\begin{equation}
\label{adm}
a^{\dag}_{n,+,\uparrow} \to t_{n,\alpha}^{\dag}a^{\dag}_{n+1,-,\sigma}
\end{equation}
which gives a second order correction to the hole energy.
Note that the symmetry of the operators $a^{\dag}$ in the admixed components
is negative.  The Hamiltonian (\ref{H})  is symmetric 
under permutation of the ladder legs. Therefore all excitations can be 
classified according to this symmetry. 
%Following standard notations we denote 
%antisymmetric excitations ($k_{\perp}=\pi$) by the index $u$ and
%symmetric ones ($k_{\perp}=0$) by the index $g$. 
The  operator $t^{\dag}$ has $(-)$-symmetry, and therefore the $a^{\dag}$
which appears together with $t^{\dag}$ must be of the $(-)$-symmetry as well.
Below in the notations of the quantum states we will omit the symmetry
index.
The energy difference corresponding to the excitation (\ref{adm}) is
\begin{equation}
\label{D}
\Delta E=\left(\epsilon_0+t+J_{\perp}-{1\over3}J+{{3J^2}\over{8J_{\perp}}}
\right)-
\left(\epsilon_0-t\right)=2t+J_{\perp}-{1\over3}J+{{3J^2}\over{8J_{\perp}}}
\end{equation}
The admixture (\ref{adm}) first of all shifts the on site energy.
Using the  wave function (\ref{in}) we find
\begin{equation}
\label{onsite}
\delta \epsilon_n=-2{{\left|\langle t_n^{\dag}a^{\dag}_{n+1}
|H_t|a^{\dag}_{n}\rangle\right|^2}\over{\Delta E}}
-2{{\left|\langle t_{n+1}^{\dag}a^{\dag}_{n}
|H_J|a^{\dag}_{n}\rangle\right|^2}\over{\Delta E}}=
-{3\over2}{{\rho^2t^2}\over{\Delta E}}
-{3\over8}{{\rho^2J^2}\over{\Delta E}}.
\end{equation}
The admixture (\ref{adm}) generates also an effective one step hopping
\begin{eqnarray}
\label{dhop}
\delta t^{(1)}&=&
-{{\langle a^{\dag}_{n+1}|H_J|t_n^{\dag}a^{\dag}_{n+1}\rangle
\langle t_n^{\dag}a^{\dag}_{n+1}
|H_t|a^{\dag}_{n}\rangle}\over{\Delta E}}
-{{\langle a^{\dag}_{n+1}|H_t|t_{n+1}^{\dag}a^{\dag}_{n}\rangle
\langle t_{n+1}^{\dag}a^{\dag}_{n}
|H_J|a^{\dag}_{n}\rangle}\over{\Delta E}}\\
&=&-{3\over4}{{\rho^2 t J}\over{\Delta E}}.\nonumber
\end{eqnarray}
There is one more contribution to the
effective one step hopping which arises due to the spin-wave quantum 
fluctuations: Let the hole first reside on  rung $n$, then a spin-wave
quantum fluctuation arises on  rungs $n+1$ and $n+2$, after that
the hole hops to  rung $n+1$ annihilating the 
spin-wave excitation $t^{\dag}_{n+1}$
and after that the spin wave excitation $t^{\dag}_{n+2}$ is annihilated
because of the interaction $H_J$ with the hole $a^{\dag}_{n+1}$.
There also exists the time reversed mechanism which gives an equal 
 contribution.
This gives
\begin{equation}
\label{dhopmu}
\delta t^{(1)}_{\mu}=
-2\mu{{\langle a^{\dag}_{n+1}|H_J|t_{n+2}^{\dag}a^{\dag}_{n+1}\rangle
\langle t_{n+2}^{\dag}a^{\dag}_{n+1}
|H_t|t^{\dag}_{n+2}t^{\dag}_{n+1}a^{\dag}_{n}\rangle}\over{\Delta E}}
=-{3\over4}{{\mu\rho t J}\over{\Delta E}}.
\end{equation}
A similar mechanism generates a two step hopping: Initially the hole is on rung
$n$, then a spin-wave quantum fluctuation arises on  rungs $n+1$ and 
$n+2$, and after that the hole annihilates this virtual excitation moving to
the rung $n+2$. Again the  time reversion of this mechanism  leads to
an 
equal contribution.
\begin{equation}
\label{t2}
t_{eff}^{(2)}=
-2\mu{{\langle a^{\dag}_{n+2}|H_t|t_{n+2}^{\dag}a^{\dag}_{n+1}\rangle
\langle t_{n+2}^{\dag}a^{\dag}_{n+1}|H_t
|t^{\dag}_{n+2}t^{\dag}_{n+1}a^{\dag}_{n}\rangle}\over{\Delta E}}
=-{3\over2}{{\mu\rho t^2}\over{\Delta E}}.
\end{equation}
Adding together (\ref{onsite}),(\ref{dhop}),(\ref{dhopmu}),
(\ref{t2}) we obtain the total second order correction to the hole energy
\begin{equation}
\label{de2}
\delta \epsilon^{(2)}_q=-{3\over{2\Delta E}}\left(\rho^2 t^2+
{1\over4}\rho^2J^2 +tJ(\rho^2+\rho\mu)\cos q + 2\rho\mu t^2 \cos 2q \right).
\end{equation}
This formula, combined with    (\ref{rung}) and (\ref{hop}) gives a reasonable
description of the hole dispersion because the energy interval $\Delta E$
is large enough. 

 The accuracy of the calculation can be improved by 
replacing the perturbation theory by a variational method.
The zeroth approximation hole wave function and the corresponding dispersion 
are of the form
\begin{eqnarray}
\label{psi0}
&&\psi_{q\uparrow}={1\over{\sqrt{N}}}\sum_n e^{iqn}a^{\dag}_{n,+,\uparrow},\\
&&\epsilon_1=\epsilon_0-t+\rho^2t\cos q.\nonumber
\end{eqnarray}
The admixed components represent the hole and the spin wave combined into 
total angular momentum ${1\over2}$ with z-projection ${1\over2}$
\begin{eqnarray}
\label{admc}
&&|+\rangle={1\over{\sqrt{N}}}\sum_n e^{iqn}
\left(t^{\dag}_n a^{\dag}_{n+1}\right)_{{1\over2}{1\over2}},\nonumber\\
&&|-\rangle={1\over{\sqrt{N}}}\sum_n e^{iqn}
\left(t^{\dag}_n a^{\dag}_{n-1}\right)_{{1\over2}{1\over2}},\\
&&\epsilon_2=\epsilon_1+\Delta E.\nonumber
\end{eqnarray}
The mixing matrix elements are
\begin{equation}
\label{me}
V=\langle+|H|\psi_{q\uparrow}\rangle=
\langle \psi_{q\uparrow}|H|-\rangle=
-{{\sqrt{3}}\over2}\rho t-{{\sqrt{3}}\over4}\rho J e^{-iq}
-{{\sqrt{3}}\over2}\mu t e^{-2iq}.
\end{equation}
Diagonalization of the matrix gives the  quasiparticle dispersion and residue
(weight of the component (\ref{psi0}) in the total wave function)
\begin{eqnarray}
\label{eZ}
&&\epsilon_q=\epsilon_1+\Delta E/2-
\sqrt{\left(\Delta E/2\right)^2+2\left|V\right|^2}\\
&&Z={1\over2}\left(1+{1\over{\sqrt{1+8\left|V\right|^2/(\Delta E)^2}}}
\right). \nonumber
\end{eqnarray}
Expansion of this solution in powers of $\left(V/\Delta E\right)^2$
gives the perturbation theory result (\ref{de2}).

We are interested in the ``isotropic'' ladder, $J_{\perp}=J$.
In this case from  perturbation theory (\ref{mu}), $\mu=-0.25$.
Comparison with more accurate computations \cite{O,Eder,ab}
shows that really $\mu=-0.3$. The difference is small, but nevertheless
in further numerical estimations we will use $\mu=-0.3$ which effectively
accounts for some higher order corrections. At $t\ll J$  Eq.(\ref{eZ})
gives
\begin{eqnarray}
\label{st}
&&\epsilon_q\approx 1.28J-0.7t+0.24t\cos q,\\
&&Z=0.85\nonumber
\end{eqnarray}
The plots of the dispersion  for 
$t/J=1,2,3$ are given in Figures 1,2,3. They agree with results of 
numerical computations \cite{Tro,Haas,Eder}.
The quasiparticle residue is close to unity.  For example
for $t/J=2$ the residue $Z\approx 0.9$ at $q=0$, 
$Z\approx 0.83$ at $q=\pi/2$, and $Z\approx 0.97$ at $q=\pi$. To avoid 
misunderstanding we note that this residue can not be directly compared
with ``photoemission spectra'' because of two reasons. First, 
 the ``photoemission'' operator is different from $a^{\dag}$ (see the 
discussion in the Ref. \cite{Eder}). The second reason is that the 
quasiparticle we consider is well defined only near bottom of the dispersion.
At small $q$ the quasiparticle energy is large and therefore it is unstable 
with respect to real emission of a spin wave which we neglect in our 
consideration.

\section{Higher spin hole states in the ladder}
It is clear that at very large $t$ the hole tends to order all the spins
ferromagnetically. This is the Nagaoka effect, as discussed in the Ref.
\cite{Tro}. The same effect produces higher spin bound states of the hole
with the  spin wave. Let us consider first the state with $S=3/2$. The simplest
ansatz for this state is
 \begin{equation}
\label{tr3}
\psi_{q}={1\over{\sqrt{N}}}\sum_n \ e^{iqn} {1\over{\sqrt{2}}}
\left(t^{\dag}_{+1,n}a^{\dag}_{n+1,+,\uparrow}+a^{\dag}_{n,+,\uparrow}
t^{\dag}_{+1,n+1}\right),
\end{equation}
where $t^{\dag}_{+1}={1\over{\sqrt{2}}}(-t_x^{\dag}-it_y^{\dag})$
corresponds to  spin projection +1. The energy corresponding to this
wave function is
\begin{equation}
\label{e3}
\epsilon_0^{(3/2)}=\epsilon_0-2t+J_{\perp}+{J\over4}+
{3\over8}{{J^2}\over{J_{\perp}}}
\end{equation}
The first term is the single 
 hole magnetic energy (\ref{e0}), the second term is 
due to hybridization, $J_{\perp}$ is the excitation energy of the triplet,
$J/4$ is due to the antiferromagnetic interaction between the hole and the
triplet, and the last term is due to the blocking of quantum fluctuations.
The state (\ref{tr3}) can propagate via intermediate virtual configurations: 
first the triplet hops and then the hole catches up.
This gives an effective hopping matrix element
\begin{equation}
\label{teff3}
t_{eff}^{(3/2)}=-{{tJ/4}\over{t+3J^2/(8J_{\perp})-J/4}}.
\end{equation}
Thus the dispersion of the spin $3/2$ state is
\begin{equation}
\label{e33}
\epsilon_q^{(3/2)}=\epsilon_0^{(3/2)}+2t_{eff}^{(3/2)}\cos q.
\end{equation}
The plots of (\ref{e33}) for $t/J=1,2,3$ are given in Figures 1,2,3.

Finally, let us consider a ferromagnetic bag with $S=5/2$
\begin{equation}
\label{tr5}
\psi={1\over{\sqrt{3}}}
\left(t^{\dag}_{+1,n}t^{\dag}_{+1,n+1}a^{\dag}_{n+2,+,\uparrow}+
t^{\dag}_{+1,n}a^{\dag}_{n+1,+,\uparrow} t^{\dag}_{+1,n+2}+
t^{\dag}_{+1,n}t^{\dag}_{+1,n+1}a^{\dag}_{n+2,+,\uparrow}\right).
\end{equation}
The energy of this state (neglecting the dispersion) is given by 
\begin{equation}
\label{e5}
\epsilon^{(5/2)}=\epsilon_0-(1+\sqrt{2})t+2J_{\perp}+{{3J}\over4}+
{3\over4}{{J^2}\over{J_{\perp}}}.
\end{equation}
For $t/J=2,3$ it is plotted in Figures 2 and 3.

Now let us  discuss our results.
In the accepted approximation the bag state with spin $S=3/2$
at arbitrary $t/J$ has  higher energy  than the  bottom of the
quasiparticle ($S=1/2$) dispersion.
Therefore the $S=3/2$ state is unstable with respect to  decay into 
a quasiparticle and a spin wave \cite{ac}. 
It exists only as a resonance in the  quasiparticle - spin wave scattering.
The $S=1/2$ quasiparticle is itself unstable in the vicinity of $q=0$.
It can decay into a quasiparticle  and a spin wave with higher $q$, or
into an $S=3/2$ state and a spin wave.

The energy of the  $S=5/2$ bound state decreases with $t$ faster than that of
the $S=1/2$ state. Therefore sooner or later the $S=5/2$ bag state
will become the lowest one and thus the true  quasiparticle of the
system.
Comparing (\ref{eZ}) with (\ref{e5}) we find the crossover point 
$t_c\approx 7J$.  As $t$ increases further, in accordance with Nagaoka's
theorem,  quasiparticles with higher spins will appear.

\section{Single hole in two coupled antiferromagnetic planes}
The Hamiltonian in this case is similar to (\ref{H}). The only difference
is that instead of the two-leg ladder we consider now two coupled planes
(we assume that both planes form square lattices). The coupling $J$ is
 the antiferromagnetic exchange in the plane,
and  $J_{\perp}$ is the antiferromagnetic exchange between the planes.
It is known that at $J_{\perp} > J_c$ ($J_c \approx 2.5J$) the system
is in a disordered dimerized spin liquid state, and at $J_{\perp}<J_c$ it
undergoes quantum phase transition to a Neel phase with long range
antiferromagnetic order (see, e.g. Refs. \cite{Chub1,K}).
In the present section we will consider  the spin liquid phase in the
regime 
 very close to the point of antiferromagnetic instability.

Let us remove one spin from the system. We will see that the most interesting
situation which is qualitatively different from the ladder arises when the
hole is static ($t=0$). Therefore, let us consider the static hole 
(impurity) problem.  
 In the first approximation, neglecting the interaction with
spin waves, the energy of the hole is
\begin{equation}
\label{e00}
\epsilon_0={3\over4}J_{\perp}+{3\over2}{{J^2}\over{J_{\perp}}}\approx 2.5J.
\end{equation}
It is similar to (\ref{e0}). The first term is due to the destroyed bond
between the planes, and the second one arises from the suppression of
 quantum fluctuations on the four links in the plane. We substitute in 
(\ref{e00}) $J_{\perp}=J_c\approx 2.5J$.
Interactions with spin waves give corrections to (\ref{e00}).
The hole-spin-wave vertex is represented by the
 $t^{\dag}{\bf S}$ term in the effective
Hamiltonian (\ref{Hh}). In  momentum representation it has the form
\begin{equation}
\label{gkq}
g_{\bf k,q}=-J \gamma_{\bf q}\left(u_{\bf q}+v_{\bf q}\right)\sigma^{\alpha}.
\end{equation}
This vertex describes the process when a hole with momentum ${\bf k}$ 
decays into a hole with momentum ${\bf k-q}$ and a spin wave with
momentum ${\bf q}$ and polarization $\alpha$, ($\alpha=x,y,z$).
The spin Pauli matrix of the hole is denoted by ${\vec \sigma}$;
$\gamma_{\bf q}={1\over 2}(\cos q_x + \cos q_y)$; $u_{\bf q}$ and $v_{\bf q}$
are the Bogoliubov parameters corresponding to the spin wave \cite{K}.
In (\ref{gkq}) we neglect the  deviation of 
 the spin wave quasiparticle residue 
from unity because it gives only a 10\% correction (see Ref. \cite{K}).
The spin wave dispersion has a  minimum at ${\bf Q}=(\pi,\pi)$. Near this point
the dispersion is
\begin{equation}
\label{om}
\omega_{\bf q}=\sqrt{\Delta^2 + c^2({\bf q-Q})^2},
\end{equation}
where $c\approx 1.9J$ \cite{Hida,K} is the  spin wave velocity.
At ${\bf q} \approx {\bf Q}$ the Bogoliubov parameters can be represented
as \cite{K}
\begin{equation}
\label{uv}
u_{\bf q}\approx v_{\bf q} \approx \sqrt{{A}\over{2\omega_{\bf q}}},
\end{equation}
where $A\approx 2.4J$.

The naive second order correction to the hole energy is given by
\begin{equation}
\label{de}
\delta \epsilon =-\sum_{\bf q}{{3g_{\bf k,q}^2}\over{\omega_{\bf q}}}
=-{{3J^2A}\over{2\pi^2}}\int{{d^2q}\over{\omega_{\bf q}^2}} \approx
-J_0  \ln{J\over{\Delta}},
\end{equation}
where the integral is calculated with logarithmic accuracy.
We have introduced here the notation
\begin{equation}
\label{J00}
J_0={{3J^2A}\over{\pi c^2}}\approx 0.63 J 
\end{equation}
It is clear that since ${J\over{\Delta}}\gg 1$, the naive formula (\ref{de})
has no meaning. However, the presence of the logarithmic divergence indicates
that the self consistent Born approximation (see e.g. Ref. 
\cite{Schmitt,Kane}) is valid in this case.
Following this approximation let us consider the quasiparticle (pole) part of
the hole Green's function
\begin{equation}
\label{G}
G(\epsilon)={{Z}\over{\epsilon-e+i \ 0}},
\end{equation}
where $e$ is the energy of the state.
This Green's function must obey Dyson's equation
\begin{equation}
\label{DE}
G(\epsilon)={1\over{\epsilon-\epsilon_0-\Sigma(\epsilon)}},
\end{equation}
where the self energy is given by
\begin{equation}
\label{SE}
\Sigma(\epsilon)=3\sum_{\bf q}g_{\bf k,q}^2 \ G(\epsilon-\omega_{\bf q})
\approx {{3J^2A}\over{2\pi^2}} \ Z \ \int {{d^2q}\over
{\omega_{\bf q}(\epsilon - e -\omega_{\bf q})}}.
\end{equation}
The calculation of this integral is straightforward and gives
\begin{equation}
\label{SE1}
\Sigma(e)=-Z J_0 \ln{J\over{\Delta}},\hspace{0.5cm}
\left({{\partial}\over{\partial\epsilon}}\Sigma(\epsilon)\right)_{\epsilon=e}
=-Z {J_0\over{\Delta}}.
\end{equation}
Finally, substituting  into (\ref{DE}) and comparing with (\ref{G})
we find
\begin{eqnarray}
\label{Ze}
e&=&\epsilon_0-\sqrt{\Delta J_0} \ \ln{J\over{\Delta}}\\
Z&=&\sqrt{{{\Delta}\over{J_0}}}.\nonumber
\end{eqnarray}
The quasiparticle residue $Z$ vanishes at $\Delta \to 0$.
It is clear that it does not happen at finite hopping, because
in this case the lower limit of integration in (\ref{de}) is $t$
and the integral remains finite at $\Delta=0$. Physically it means
that due to the finite speed of propagation the hole has no time for
dressing by an infinite number of spin waves. 

Thus, the static impurity is strongly dressed by  spin waves.
The question arises: is this cloud dilute enough to justify
the spin-wave approach we have applied, or is it dense in which case the above
consideration has no quantitative meaning because we do not take into
account  the  interaction of the spin waves between themselves?
The size of the cloud is $r \sim c/\Delta$ (in units of the  lattice spacing),
and the effective number of  spin waves is $N_{sw}\sim 1/Z =\sqrt{J_0/\Delta}$.
Therefore the density of spin waves $n\sim N_{sw}/r^2 \sim 
{{\sqrt{J_0}}\over{c^2}}\Delta^{3/2}$ remains low and hence the dilute 
gas approximation is valid.

\section{Conclusions}
Using variational approach and perturbation theory we have found 
analytically the dispersion of a single hole with spin $S=1/2$ 
 in the $t-J$ ladder.
The dispersion is in an agreement with previous computations.
We demonstrate also that there are higher spin ($S=3/2,5/2,...$)
bound states of the hole with the  magnetic excitations.
At large $t/J$ these bound states lie below the  $S=1/2$ state meaning that
they become the real quasiparticles of the system.
We have estimated the critical value of $t/J$ where the crossover to the higher
spin quasiparticles take place.

In spite of the complex spectrum, the structure of  every state
on the ladder is relatively simple. This means that the number of significant
components in the wave function remains of the order of unity. 
In the second part of the present work we have considered the opposite
situation when the wave function has a very large number of significant
components. This is realized for a static hole in two antiferromagnetic planes 
close to the point of antiferromagnetic instability.
We demonstrate that the self consistent Born approximation is
justified in this case. Using this method we show that
as the point of instability is approached the number of components in the hole
wave function  approaches  infinity and the quasiparticle residue
vanishes.

\section{Acknowledgments}
I am very grateful to V. N. Kotov,  M. Yu. Kuchiev and R. Eder   
for stimulating discussions.

\begin{figure}
\caption
{Dispersions of the $S=1/2$ quasiparticle (solid line) and the $S=3/2$ bound
state (dashed line); $t/J=1$.}
\label{fig.1}
\end{figure}

\begin{figure}
\caption
{Dispersions of the $S=1/2$ quasiparticle (solid line), $S=3/2$ bound
state (dashed line), and $S=5/2$ bound state (dashed-dotted line);
$t/J=2$.}
\label{fig.2}
\end{figure}
 
\begin{figure}
\caption
{Dispersions of the $S=1/2$ state (solid line), $S=3/2$  state (dashed line), 
and $S=5/2$ bound state (dashed-dotted line); $t/J=3$.}
\label{fig.3}
\end{figure}


\begin{references}
\bibitem{many} To derive the  mapping for the case of many holes
  one has to include
several additional terms which reflect hole-hole interactions.
\bibitem{ab}
This is the perturbation theory formula valid, strictly speaking, at 
$J_{\perp} \gg J$. However  it works well 
even at $J_{\perp}\sim J$. For example at $J_{\perp}=J$, the
perturbation theory gives  $\delta E_0=-0.375J$, and according to exact 
numerical  calculations \cite{O} $\delta E_0=-0.404J$.
\bibitem{ac} The spin-wave spectrum can be found in Ref. \cite{O}.
At $J_{\perp}=J$ the minimum energy of the spin wave is 
$\omega(q=\pi) \approx J/2$. This is the decay threshold.
\end{references}
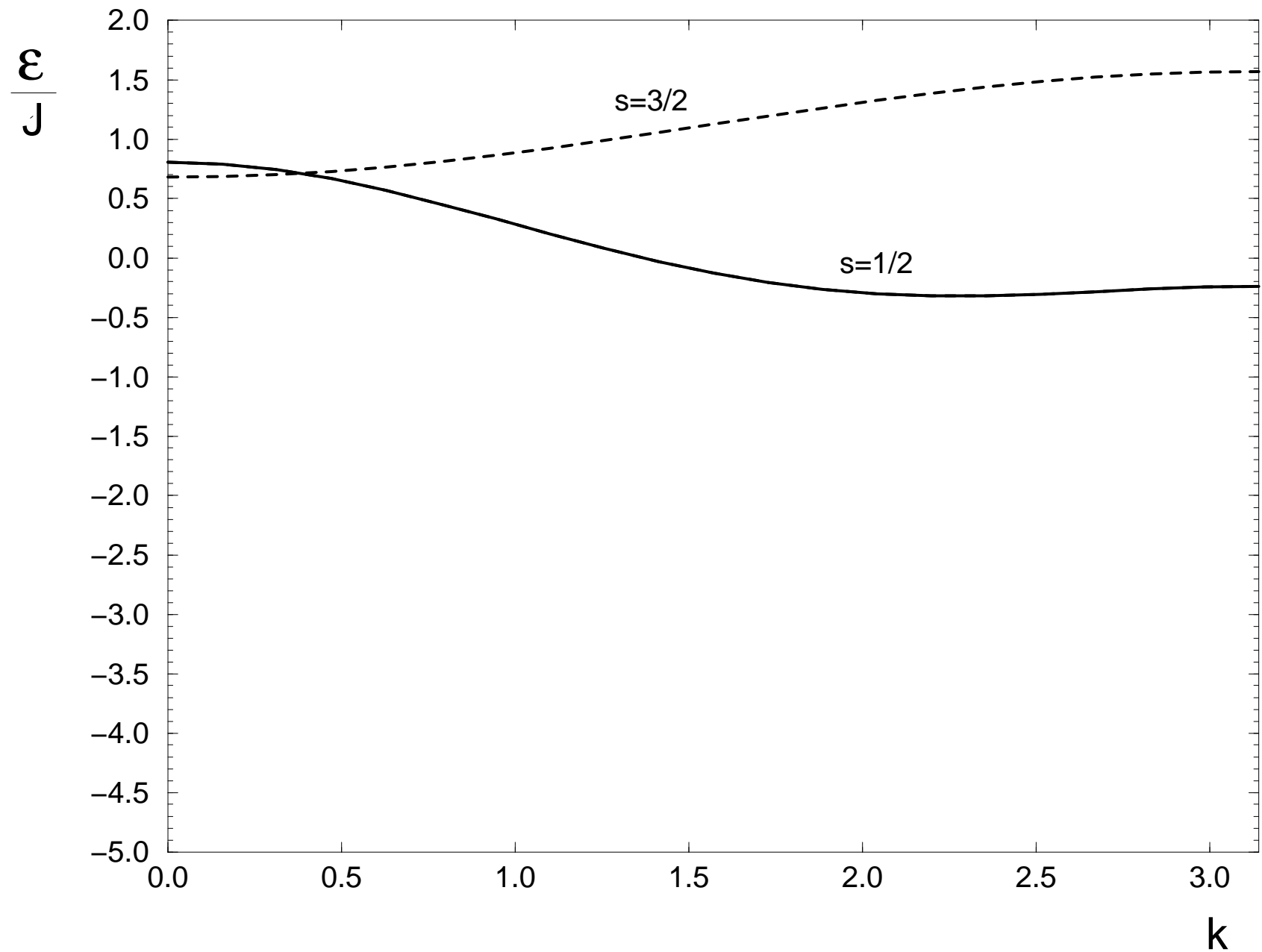
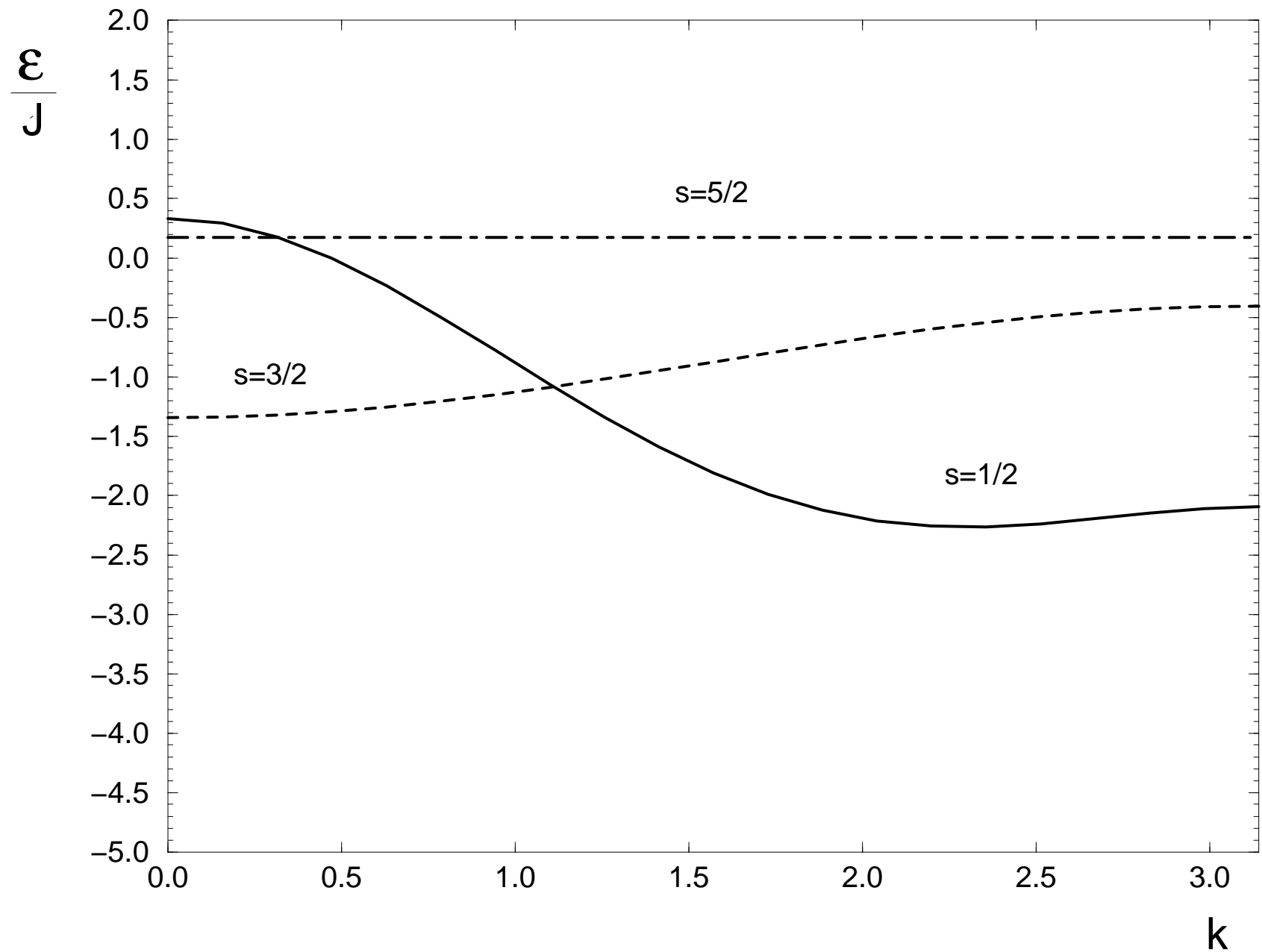
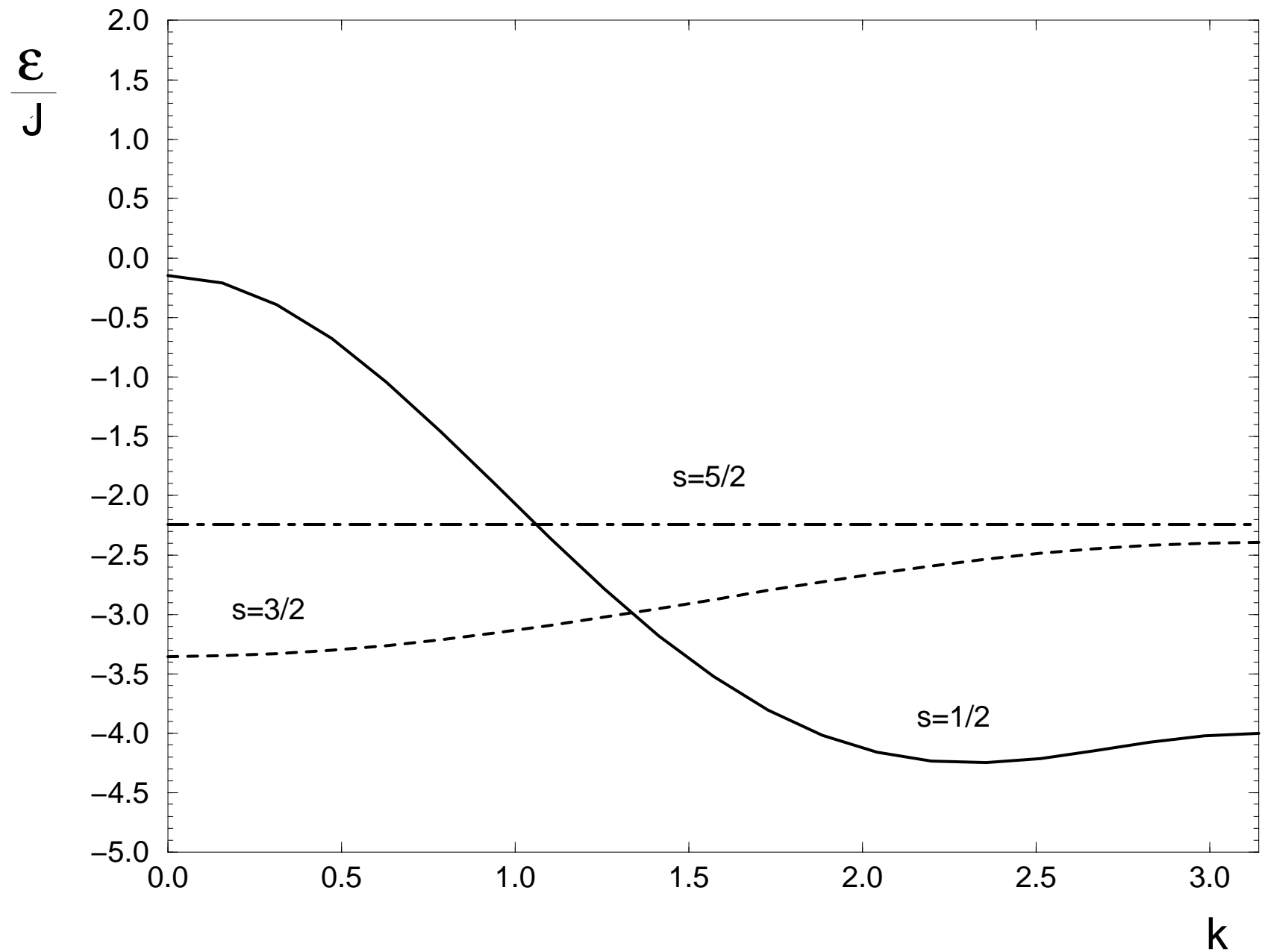
\end{document}